\documentclass{vak2024}
\usepackage[utf8]{inputenc}
\usepackage{url}
\usepackage{hyperref}

\begin{document}
\title{Merging of spiral galaxies: observations and modeling}
\titlerunning{Merging of spiral galaxies}  
\author{A.~Khoperskov\inst{1}\and S.~Khrapov\inst{1}\and D.~Sirotin\inst{1}\and A.~Zasov\inst{2,3}}
\authorrunning{Khoperskov et al.} 
	%
	%
\institute{Volgograd State University, Volgograd, 400062, Russia
\and Sternberg Astronomical Institute, Moscow M.V. Lomonosov State University, Universitetskij pr., 13, Moscow, 119234, Russia
\and Faculty of Physics, Moscow M.V. Lomonosov State University, Leninskie gory 1, Moscow, 119991, Russia
		}
\abstract{
We present a study of the dynamics of multi-component models of spiral galaxies at various stages of grand merging. Numerical models include a self-consistent account of the dynamics of collisionless stellar subsystems and N-body dark matter, as well as gaseous components. Calculation of gas heating and cooling processes allows us to consider a wide temperature range from 80 to 100 thousand degrees. Using the Smoothed-particle hydrodynamics method to solve hydrodynamic equations makes it possible to track the evolution of the gas of each galaxy, calculating the content of gas components of each object in the process of complex interchange of matter. The gravitational interaction is determined in a direct way by summing the contributions according to Newton’s law, which minimizes the modeling error. This requires significant computing resources using graphics accelerators  on
hybrid computing platform CPU + multi-GPUs. The study aims to reconcile theoretical models with the morphology and kinematics of a number of observed systems. In particular, Taffy-type objects are considered, where two galaxies are connected by a gas bridge with a characteristic small-scale gas structure after the disks pass through each other in an approximately flat orientation. Examples of such systems are the observed pairs of galaxies UGC 12914/UGC 12915, NGC 4490/NGC 4485, UGC813/UGC816 etc.
 \\
{\bf To cite:} Khoperskov A., Khrapov S., Sirotin D., Zasov A. Merging of spiral galaxies: observations and modeling // Modern astronomy: from the Early Universe to exoplanets and black holes. Special Astrophysical Observatory of the Russian Academy of Sciences . 2024. pp. 149-154.
\doi{10.26119/VAK2024.025}
}

\maketitle

\section{Introduction}

Galactic interactions are the most important factor in the evolution of significant fraction of these stellar systems. The most large-scale events are associated with major mergers, leading to formation of complex tidal structures, a significant restructuring of star formation history, determining the morpho\-logy and kinematics in the process of merging of galaxies \citep{1, 2, 3}.

\begin{figure}
\centerline{\includegraphics[width=0.8\textwidth]{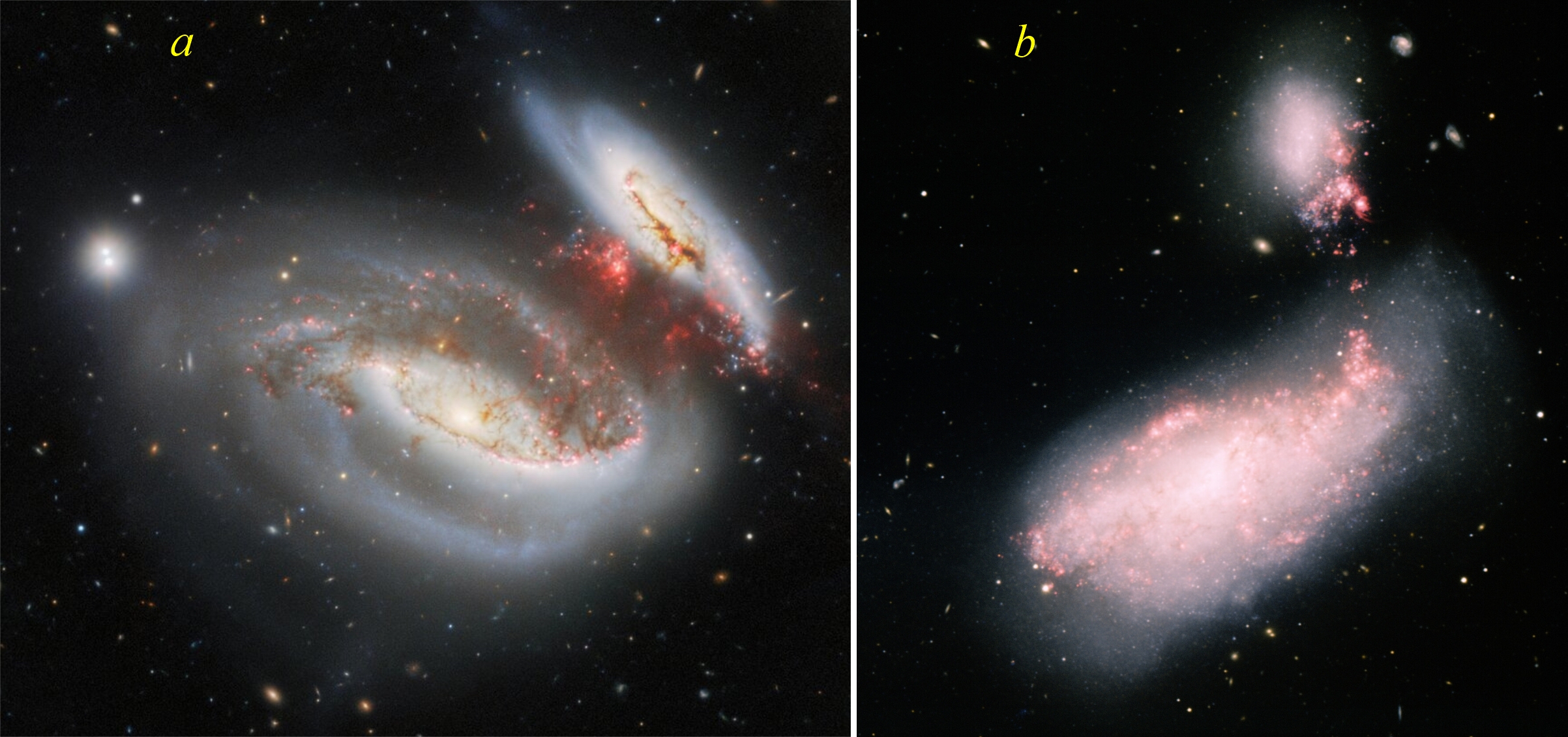}
 }
\caption{Images of Taffy galaxies: $a$ --- UGC 12914 (bottom) and UGC 12915 (top); $b$ --- NGC4490 and NGC4485 (Gemini North, https://noirlab.edu)
 }
\label{fig:UGC12914lef-UGC12915right}
\end{figure}

A very curious and rare class of interacting objects is the systems of   recently collided disky galaxies, a study of which began with the pair UGC 12914/12915 called Taffy \citep{1}. A near head-on collision occurred approximately 25 million years ago, and a gas bridge with a peculiar morphology and physical conditions of matter is clearly visible between the disks of galaxies (Fig. \ref{fig:UGC12914lef-UGC12915right}).

The goal of this paper is to construct  the numerical dynamic models of a nearly face-on collision of two S-galaxies that reproduce the main properties of the observed of Taffy-like objects. The model of each galaxy includes stellar and gas disks immersed in a live dark halo. We investigate the influence of various parameters of the galactic pair on the properties of the gas bridge observed in the in the Taffy and Taffy-like systems.

\section{Features of Taffy pairs and their modeling }

The pair UGC 12914/12915 is a classic example of Taffy galaxies, which was highlighted in the work of \cite{1}. The presence of thin filamentary gas structures between two S-galaxies is a distinctive feature of Taffy. This observed feature is formed as a result of the recent passage of the disks through each other.

Some properties of UGC 12914/12915 are as follows \citep{1, 2, 3, 4}.
 1) An almost head-on collision of two counter-rotating disks occurred several tens of millions of years ago.
 2) There is a massive multiphase gaseous bridge containing filamentary structures.
 3) The gaseous bridge includes warm molecular hydrogen.
 4) Star formation in the bridge is low, if any, evidently suppressed by the strong  gas turbulence. 
 5) The soft X-ray emission from the bridge is associated with hot gas with a mass in the range of $(0.8-1.3)\cdot 10^8 M_\odot$, which is about 1\% of the total mass of the gas between the disks \citep{5}.  

The pair NGC 4490 and NGC 4485 are also Taffy-type (Fig. \ref{fig:UGC12914lef-UGC12915right}), although the impact geometry was apparently different from the UGC 12914/12915 system. 
Note that warm molecular hydrogen has been detected in outflows of gas in M82. Such superwinds are a fairly common phenomenon, the structure of which is also filamentary.

\begin{figure}[!t]
\centerline{\includegraphics[width=0.995\textwidth]{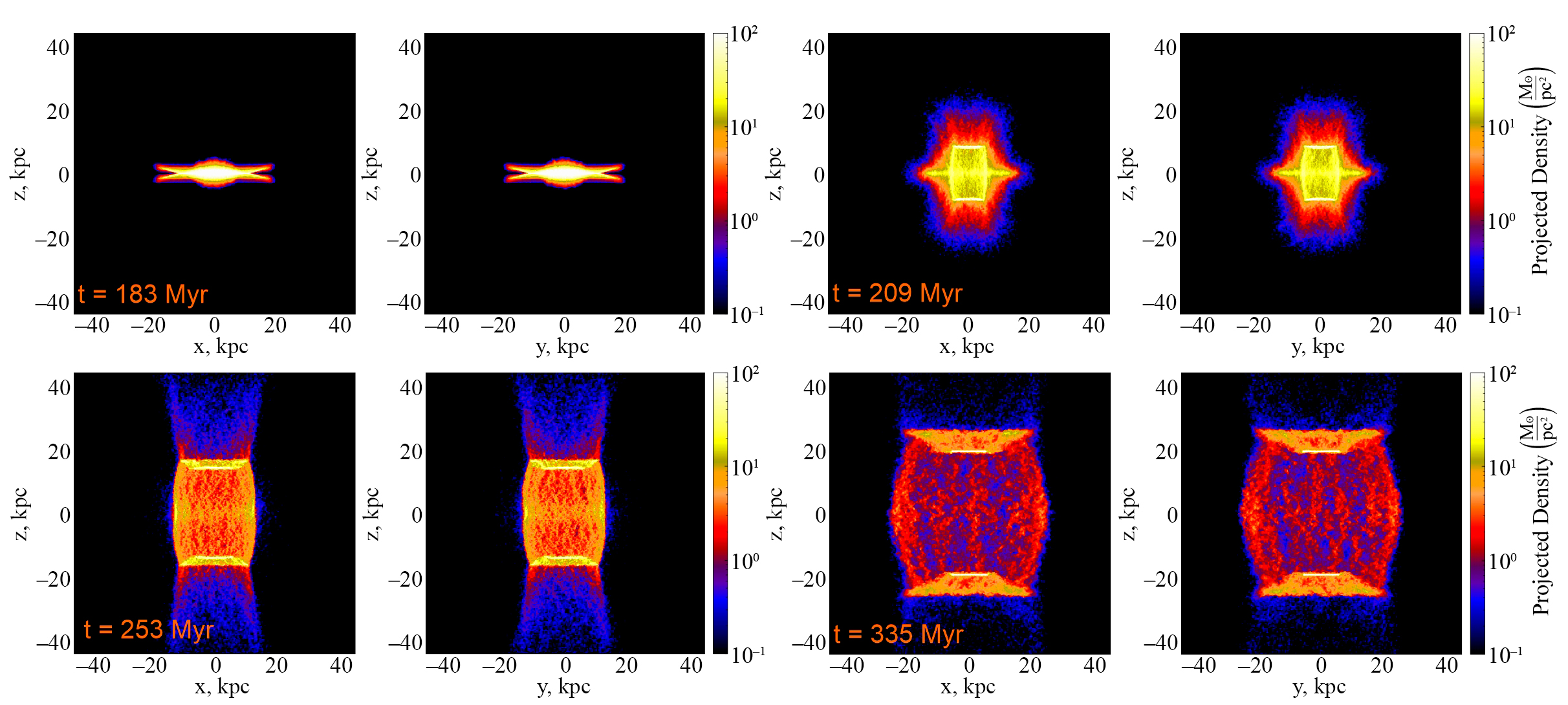}}
\caption{
An example of a head-on collision simulation of a Taffy-type pair with retrograde rotation of the disks. }
\label{fig:Evolution-collision-Taffy}
\end{figure}

\section{Numerical simulation of Taffy galaxies}

We present preliminary results from simulations of two colliding galaxies, each contain\-ing a stellar disk, a gas disk, and a dark halo. The basic model consists of identical galaxies with masses of stars $M_d = 3.72 \times 10^{10} M_\odot$, gas $M_g = 3.72 \times 10^{9} M_\odot$, dark matter $M_h = 6.02 \times 10^{10} M_\odot$ (within the optical radius $R_{opt} = 9$ kpk). The total mass of the simulated dark halo is $ 24.92 \times 10^{10} M_\odot$ in each galaxy.

Two galaxies possessing stellar disks with masses $M_d^{(1)}$ and $M_d^{(2)}$ and dark halos with masses $M_h^{(1)}$ and $M_h^{(2)}$ are simulated by collisionless N-body particles. Gas dynamics are calculated by the smoothed-particle hydrodynamics (SPH) method. Gravitational interaction is calculated by direct summation of contributions from all particles (N-body + SPH-particles) \citep{6, 7, 8}.

Fig. \ref{fig:Evolution-collision-Taffy} shows the distribution of gas density along the line of sight during a head-on collision of two identical galaxies. The time $t=183$ Myr corresponds to the first contact of gas disks. The following frames give a picture of the formation of a gas bridge between galaxies. Filamentary structures are a characteristic feature of dynamics of a gas swept out by the impact.
The bridge region does not contain stars.
The considered flat-type impact distorts the stellar disk significantly less than the gas disk, which undergoes the dramatic changes.

\begin{figure}
\centerline{\includegraphics[width=0.6\textwidth]{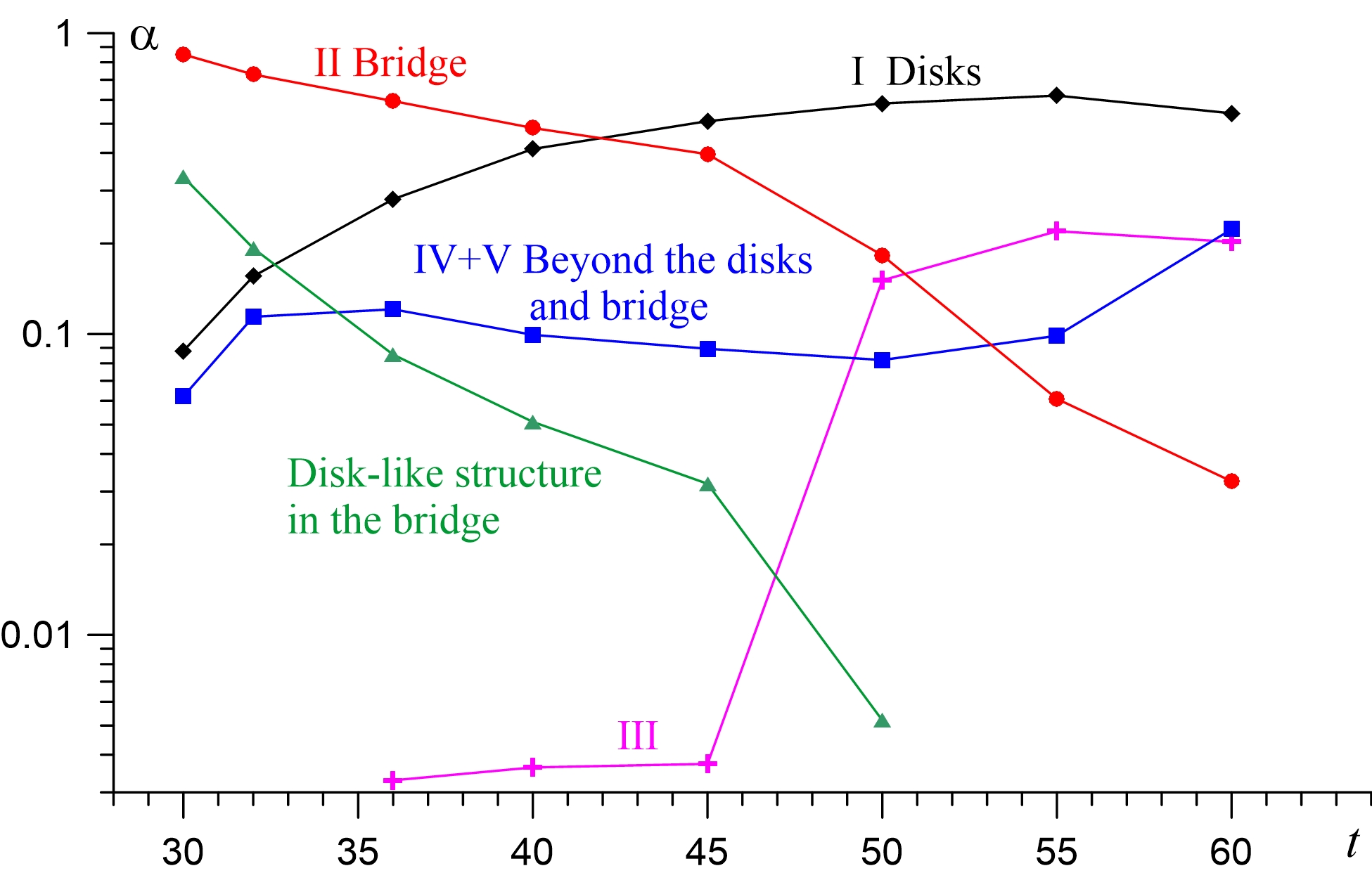}}
\caption{
Dependence of gas mass in the different regions of the interacting system during the evolution process, $t=1\rightarrow  63.2$\,Myr. 
 }
\label{fig:Mass-gas-Bridge}
\end{figure}

The following regions of the Taffy system can be conditionally distinguished (See Fig.~\ref{fig:Evolution-collision-Taffy}): I~--- gas inside stellar disks; II --- gas bridge between galactic disks, determined by the conditions $z^{(disk)}_1 < z < z^{(disk)}_2$, $r< R_{gas}^{(disk)}$; III --- outer zone relative to the bridge between galaxies, $z^{(disk)}_1 < z < z^{(disk)}_2$, $r> R_{gas}^{(disk)}$, IV and V --- remaining zones outside the galactic disks, $z > z^{(disk)}_1$ (IV), $ z < z^{(disk)}_2$ (V). Fig.~\ref{fig:Mass-gas-Bridge} shows the changes in gas abundance over time for these four regions.

\begin{figure}[!t]
\centerline{\includegraphics[width=0.60\textwidth]{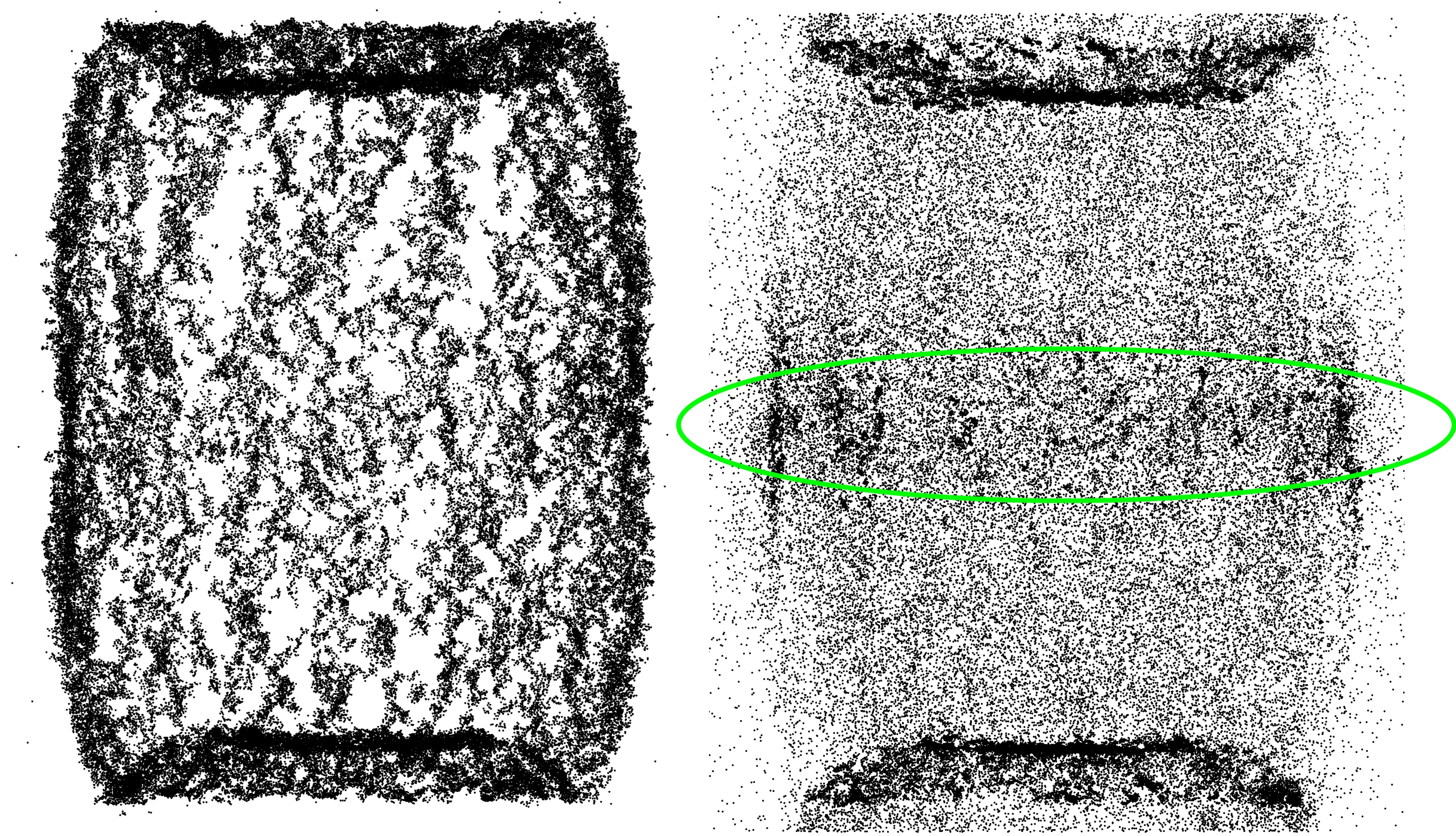}
 } 
\caption{
Structure of warm gas (left) and neutral hydrogen (right) between galaxies 70 million years after the first impact. The disk-like structure in the bridge area is highlighted by green line.
 }
\label{fig:Structure-disc-after}
\end{figure}

Fig. \ref{fig:Structure-disc-after} shows the distribution of gas between the disks depending on the gas temperature for a symmetrical head-on impact. Hot gas at a temperature of $3000-10000$ K is distributed fairly uniformly. There are only residues of increased concentration near the plane of symmetry ($z=0$), which is associated with the position of the initial impact. Relatively cool gas ($100-500$ K) forms inhomogeneous filamentary structures.

The cold gas forms a shell structure in contrast to the hotter component, which occupies the entire volume of the bridge. This feature appears due to complete symmetry of the system when two identical galaxies collided. Simulating the impact of dissimilar galaxies or the lack of geometric symmetry of the initial interaction critically changes the properties of the gas bridge.

The overall result is that the gas bridge can contain up to 90 percent of the total gas mass immediately after the impact. This mass decreases down to 30 percent as the disks move away from each other. The bridge loses gas due to gas deposition on the disks of galaxies and gas expansion in the bridge region. We calculate the gas fractions in each region from each galaxy separately. For example, $\alpha^{(I)}_1$ is the fraction of the gas mass in region I from the first galaxy. If the initial galaxies are not identical or the collision geometry is not symmetric, then  $\alpha^{(II,III)}_1 \ne 1/2$. Computational experiments show fairly strong dependencies of the parameters $\alpha$ on the collision conditions. We also calculate the enrichment of the first galaxy in the gas of the second one and vice versa. These characteristics $\alpha^{(I)}_{12}$ and $\alpha^{(I)}_{21}$ are quite sensitive to the initial collision conditions. The green line in Fig. \ref{fig:Mass-gas-Bridge} shows the presence of a massive disk-like structure in the bridge region (more than 30 percent of the total gas) immediately after the impact in the $z=0$ plane for identical galaxies (See Fig.~\ref{fig:Structure-disc-after}). However, this gas dissipates quickly. Fig.~\ref{fig:Structure-disc-after} (right) shows the remaining remnants of this gas 70 million years after the first impact.

\section{Summary}

The results allow to account for the presence and a structure of inhomogeneous multi-component media observed in  Taffy-type galaxies.
We have performed more than 30 computational experiments to study the structures formed when disk galaxies collide  almost face-on.
The main focus of the study is to determine the properties of gas in the bridge region between the disks of galaxies. 
Our results show that the spatial structure of  gas and its thermodynamic characteristics are very sensitive to the impact geometry and parameters of the parent galaxies.
Our numerical models allow us to distinguish gas components that form the spatially distinct subsystems with different temperatures within the range of  $50\div 2\cdot 10^5$\,K.

\section*{Funding} 
This work supported by the Russian Science Foundation~(grant no. 23-71-00016,  https://rscf.ru/project/23-71-00016/). The research also relied on the shared research facilities of the HPC computing resources at Lomonosov Moscow State University.


\end{document}